\documentclass{article}
\usepackage{spconf,amsmath,graphicx}
\usepackage{dcolumn}

\usepackage{enumitem}
\usepackage{setspace}
\usepackage{booktabs} 
\usepackage{color} 
\usepackage{multirow} 
\usepackage{graphicx}
\usepackage{subcaption}
\usepackage{amssymb}
\usepackage{enumitem}
\usepackage{cite}

\title{improving noise robustness of automatic speech recognition via parallel data and teacher-student learning}

\makeatletter
\def\@name{\emph{Ladislav Mo\v{s}ner}$^1$$^*$\thanks{$^*$ Ladislav Mosner performed the work while he was a research intern at Amazon.}, \emph{Minhua Wu}$^2$, \emph{Anirudh Raju}$^2$, \emph{Sree Hari Krishnan Parthasarathi}$^2$,\\ \emph{Kenichi Kumatani}$^2$, \emph{Shiva Sundaram}$^2$,  \emph{Roland Maas}$^2$, \emph{Bj{\"o}rn Hoffmeister}$^2$}
\address{\\$^1$Brno University of Technology, Faculty of Information Technology, IT4I, Czechia Czechia\\   $^{2}$Amazon.com, Inc. USA\\ \texttt{\small imosner@fit.vutbr.cz, wuminhua@amazon.com}}
\makeatother
\begin{document}
\maketitle
\begin{abstract}
For real-world speech recognition applications, noise robustness is still a challenge. In this work, we adopt the teacher-student (T/S) learning technique using a parallel clean and noisy corpus for improving automatic speech recognition (ASR) performance under multimedia noise. On top of that, we apply a logits selection method which only preserves the $k$ highest values to prevent wrong emphasis of knowledge from the teacher and to reduce bandwidth needed for transferring data. We incorporate up to 8000 hours of untranscribed data for training and present our results on sequence trained models apart from cross entropy trained ones. The best sequence trained student model yields relative word error rate (WER) reductions of approximately 10.1\%, 28.7\% and 19.6\% on our clean, simulated noisy and real test sets respectively comparing to a sequence trained teacher.
\end{abstract}
\begin{keywords}
automatic speech recognition, noise robustness, teacher-student training, domain adaptation
\end{keywords}
\section{Introduction}
\label{sec:intro}

With the exponential growth of big data and computing power, automatic speech recognition (ASR) technology has been successfully used in many applications. People can do voice search using mobile devices. They can also interact with smart home devices such as Amazon Echo or Google Home through distant speech recognition \cite{kumatani2012microphone}\cite{wu2019multichannelAM} for entertainment, shopping or other personal assistance. For such real-world applications, noise robustness is important since the device needs to work well under various acoustic environments, and it still remains to be a challenging task \cite{li2014overview, king2017robust}. Although large-vocabulary speech recognition is of high accuracy by applying deep neural networks \cite{sainath2011making,dahl2012context,hinton2012deep}, it requires thousands of hours of transcribed data which is time-consuming and expensive to collect, and its performance under noisy environment may still suffer. 
Considerable efforts have been made to improve noise robustness by applying algorithms in the front-end feature domain \cite{boll1979suppression, atal1974effectiveness,acero1990acoustical,hermansky1994rasta,hermansky1990perceptual,Macho2002EvaluationOA,cui2002evaluation} or in the back-end model \cite{gales1996robust,moreno1996speech,leggetter1995maximum,wu2002supervised,he2003minimum,yu2009unsupervised}. Another natural way to deal with noise in the acoustic environment is to use multi-style training \cite{lippmann1987multi}, which trains the acoustic model with noisy speech data. All these approaches require \textit{supervision} where the speech data is manually transcribed.

In order to improve noise robustness of the distant speech recognition system in an \textit{unsupervised} mode, it is desirable to scale multi-style training with an even larger training dataset. However, acquiring manually transcribed data for noisy speech can be slow and expensive. In this work, we explore the technique of teacher-student (T/S) learning using a parallel corpus of clean and noisy data. We focus on improving the ASR performance under multimedia noise which is commonly present at home.

T/S learning was at first explored in speech community \cite{li2014learning} \cite{hinton2015distilling} to distill the knowledge from bigger models to a smaller one, and was successfully applied in the areas of ASR \cite{chebotar2016distilling,lu2017knowledge} and keyword spotting \cite{tucker2016model} afterwards. Instead of knowledge distillation, we adopt the T/S learning for domain adaptation which was proposed in \cite{Li:Domain-adapt} to build an ASR system performing more robustly under multimedia noise. On top of this system, we apply logits selection keeping only the $k$ highest values and experiment it with multiple settings of temperature $T$. This method was proposed in \cite{hari2019onemillionAM} to optimize storage and to parallelize the target generation for teacher-student training, and we find that it even helps improve performance of the adapted student model since it prevents over-emphasizing wrong senones by the teacher. Because the T/S learning technique applied in this work does not require transcribed data, we also explore how much the system performance can be further improved by gradually incorporating more training recordings. Finally, we study the effects of doing sequence training on top of the T/S learning for domain adaptation, which was not reported in \cite{Li:Domain-adapt}.
\vspace{-1mm}

\section{Method}
\begin{figure}[h]
\centering
  \includegraphics[width=0.72\linewidth]{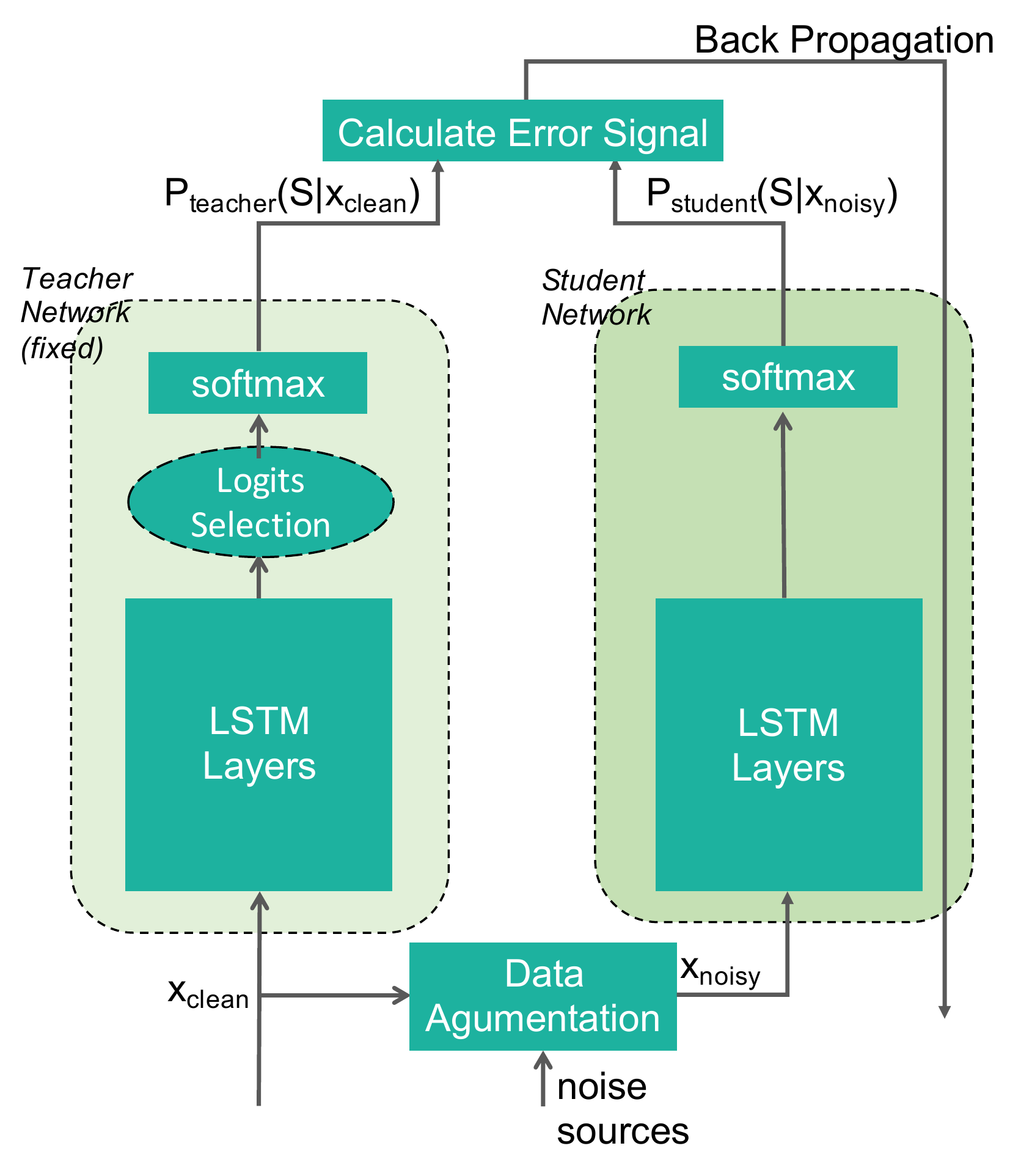}
  \caption{\small{\textit{The flow chart of teacher-student learning using parallel data for improving noise robustness of ASR.}}}
  \label{fig:setup}
  \vspace{-4mm}
\end{figure}

\subsection{Teacher-student training for domain adaptation}
\label{sec:ts-training}
In order to apply T/S training for domain adaptation (to adapt from clean teacher to noisy student domain in our use case), parallel clean and noisy corpus is needed and it could be generated by artificially adding noise on top of the clean data, which will be described in detail in section \ref{sec:train-dataset}. As shown in Figure \ref{fig:setup}, a teacher network outputs discrete probability distribution $P_{teacher}(S | x_{clean})$ over senones $S$ given a clean feature vector $x_{clean}$, and a student network estimates probabilities of senones $P_{student}(S|x_{noisy})$ given a noisy feature vector $x_{noisy}$.  The objective is to minimize the Kullback-Leibler (KL) divergence between these two distributions in order to make behavior of the student network in the target domain approach that of the teacher network in the source domain.

Since the teacher parameters are fixed when training the student model, minimizing the KL divergence is equivalent to minimizing the cross entropy between output distributions from the teacher and the student, where the teacher's output distribution (``soft targets") are considered as the ground truth. Therefore, student training relies completely on teacher outputs and no transcriptions are required.

\subsection{Logits selection}
\label{sec:logit-encoding}
The discrete probability distribution over senones at each time frame from the teacher will guide the learning of the student model. Since there are usually thousands of senones for a typical ASR system, a large number of values need to be loaded when processing just one frame. Moreover, the majority of the output probability mass is usually covered by just a few outputs, while the rest of output values are very small and noisy, which might confuse the student model instead. In this case, we adopt the logits selection method \cite{hari2019onemillionAM} with which only $k$ highest logit values are preserved so that only the most reliable information from the teacher is preserved to train the student. At the  
same time, this method also reduces the bandwidth of transferring frames of soft targets for training. Let $z_i$ be an activation before the softmax (logit), $N$ be number of classified senones. Without any selection, the output probability of senone $s_i$ is computed as follows:
\begin{equation}\label{eq:posteriorNoSelection}\small
P_{teacher}(S=s_i | x_{clean}) = q_i = \frac{ \exp(z_i/T) }{ \sum_{j=1}^{N} \exp(z_j/T) },
\end{equation}
where T is the temperature controlling softness of the distribution\cite{hinton2015distilling}.

When considering only $k$ highest logit values whose indices lie in the set $\mathbb{K}$, the $k$ highest probabilities of senones are preserved with an emphasizing factor $\mathbf{A}$ while the rest are suppressed to be zero.
\begin{equation} \label{eq:posteriorKLogits}\small
q'_i =
    \left \lbrace \begin{array}{ll}
        \frac{ \exp(z_i/T) }{ \sum_{j\in\mathbb{K}} \exp(z_j/T) } = \mathbf{A} q_i & \text{if~~} i \in \mathbb{K}, \\
        0 & \text{otherwise.} \\
    \end{array} \right.
\end{equation}
\begin{equation}\label{eq:emphasizeFactorA}\small
\mathbf{A} = \frac{\sum_{j=1}^{N} \exp(z_j/T)}{\sum_{j\in\mathbb{K}} \exp(z_j/T)}
\end{equation}

With this selection method, we are able to dramatically reduce storage space needed for soft targets and also I/O operations during training since $k$ is significantly small compared with $N$. In fact, we notice that this method even helps to improve performance of the student model since it is boosting confidence of the teacher model and suppressing the confusing part as indicated in Equations (\ref{eq:posteriorKLogits}) and (\ref{eq:emphasizeFactorA}). 

\vspace{-1mm}
\section{Experimental setup}
\label{sec:exp-setup}
\subsection{Model architecture}

Both teacher and student have the same architecture for all the experiments. The neural network consists of three LSTM \cite{sak2014long} layers each of which comprises 512 units. The last LSTM layer is followed by a fully connected layer outputting probability distribution over 3,010 senones. 64-dimensional feature vectors of Log Mel-Filter-Bank Energies (LFBE)  are used as inputs to the network. There is no frame stacking, and the output HMM state label is delayed by 3 frames. We run decoding in a single-pass framework using a smoothed 4-gram language model (LM) trained on both internal and external data sources. The total text used to train the LM is over billion words.\footnote{Our experimental ASR system does not reflect the performance of the production Alexa system.}

\subsection{Parallel training datasets}
\label{sec:train-dataset}
The original clean corpus consists of approximately 8,000 hours of beamformed speech recordings. However, only 800 hours are transcribed and thus are used to build the teacher model with supervised training. We perform data simulation \cite{raju2018augmentation} to obtain the parallel noisy training data. Since our primary aim is to deal with multimedia noise in room conditions, we collected a corpus consisting especially of music samples and acoustic video content. For every utterance, one to three additional noises are randomly selected. The notion of sound propagation in enclosures is obtained by means of room simulation. The image method is used to acquire artificial room impulse responses \cite{Allen:Image-method, pyroomacoustics}. Reverberation times are uniformly drawn from the interval of $(500, 900)$~ms. The mixture of noises is combined with the clean signal at signal-to-noise ratio (SNR) ranging from 0 to 30 dB.

\subsection{Test datasets}
In order to evaluate the performance of our acoustic models, we used three test datasets.
\begin{description}[style=multiline,leftmargin=1.8cm,font=\normalfont]
\item[``clean":] It is of similar domain to that of the clean training data ($\sim$ 41k utterances).

\item[``noisy":] It is derived from the ``clean" dataset by the same simulation method described in section \ref{sec:train-dataset}, but different multimedia noise sources are selected and reverberation times are drawn from the different interval of (520, 920)~ms.

\item[``realistic":]  It is collected in a real room ($\sim$ 2k utterances). Clean speech and multimedia noises are played by loudspeakers and recorded by multiple microphones varying in their positions.
\end{description}

\section{Results}
\label{sec:results}

\subsection{Multi-condition versus teacher student training}
\label{sec:seq-train}
A teacher acoustic model is trained using the transcribed 800-hour clean corpus and treated as the baseline. The conventional multi-condition \cite{lippmann1987multi} trained acoustic model is also built using the transcribed portion of the noisy training dataset. As displayed in Table \ref{tab:temperatures}, the multi-condition trained acoustic model outperforms the teacher by a significant margin on the noisy test set.  At the same time, its performance on the clean test set is comparable to the performance of the baseline.

When training the student acoustic model, we at first initialize it using the baseline teacher model. During the training process, the cross entropy between $P_{student}(S | x_{noisy})$ and $P_{teacher}(S | x_{clean})$ are being minimized.

When the temperature is set to $T=1$ (standard softmax), we observe improvements in performance over the multi-condition trained model on all the datasets. Matching the student output probability distribution with that of the teacher enhances generalization ability of the student and we see an average relative reduction of about 1.9\% in WER even on the clean test dataset. The increased temperature does not seem to help achieve further improvements but leads to worse performance. A temperature of $T=5$ even results in higher WER in comparison with the multi-condition trained model, which indicates that a flatter output distribution from the teacher may result in confusions for the student to learn effectively.
\begin{table}[h]
    \small
    \caption{\small \textit{{Comparison of multi-condition and teacher-student training with different temperatures (no logits selection). Results are reported in relative word error rate reduction (WERR) [\%]. Minus sign indicates improvement.}}}
    \centering
    \begin{tabular}{l r r r}
        \toprule
        Acoustic model&\begin{tabular}{@{}r@{}} Clean \\test set \end{tabular} & \begin{tabular}{@{}r@{}} Noisy \\test set \end{tabular} & \begin{tabular}{@{}r@{}} Realistic \\test set \end{tabular} \\
        \midrule
        Baseline/teacher (clean 800h) & 0.00 & 0.00 & 0.00 \\
        Multi-condition (noisy 800h) & 0.69 & -15.20 & -4.26\\
        Student, $T=1$ (parallel 800h) &  \textbf{-1.93} &  \textbf{-17.20} & \textbf{-6.82} \\
        Student, $T=2$ (parallel 800h) & -0.46 & -16.43 & -6.35 \\
        Student, $T=5$ (parallel 800h) & 1.16 & -13.50 & -2.20 \\
        \bottomrule
    \end{tabular}
    \label{tab:temperatures}
    \vspace{-6mm}
\end{table}

\subsection{Number of candidate logits and temperature}
\label{sec:encoding-and-temperature}
To prevent needless emphasis of wrong senones from the teacher and reduce the bandwidth needed for transferring soft targets, we explore the logits selection approach explained in section \ref{sec:logit-encoding}. In our experiments, we slightly changed the computation of output probabilities previously defined by Equations (\ref{eq:posteriorKLogits}) and (\ref{eq:emphasizeFactorA}). This modification is performed for the convenience of training but does not break the general idea. Instead of assigning zero probability to the non-$k$-best senones, we assign a sufficiently high negative constant $C$ to the corresponding logits. The output probability now becomes

\begin{equation}\label{eq:posteriorKLogitsC}\small
\tilde{q'_i} =
    \left \lbrace \begin{array}{ll}
        \frac{ \exp(z_i/T) }{ (N-k)\exp(C/T) + \sum_{j \in\mathbb{K}} \exp(z_j/T) } = \tilde{\mathbf{A}}q_i & \text{if~~} i \in\mathbb{K}, \\
        \frac{ \exp(C/T) }{ (N-k)\exp(C/T) + \sum_{j \in\mathbb{K}} \exp(z_j/T) } \approx 0 & \text{otherwise.} \\
    \end{array} \right.
\end{equation}

\begin{equation}\label{eq:emphasizeFactorAwithC}\small
\tilde{\mathbf{A}} = \frac{\sum_{j=1}^{N} \exp(z_j/T)}{(N-k)\exp(C/T) + \sum_{ j \in \mathbb{K}} \exp(z_j/T)}
\end{equation}

\begin{figure*}[!h]
\begin{center}
\begin{minipage}[t]{0.26\linewidth}
  \centering
    \includegraphics[width=\linewidth]{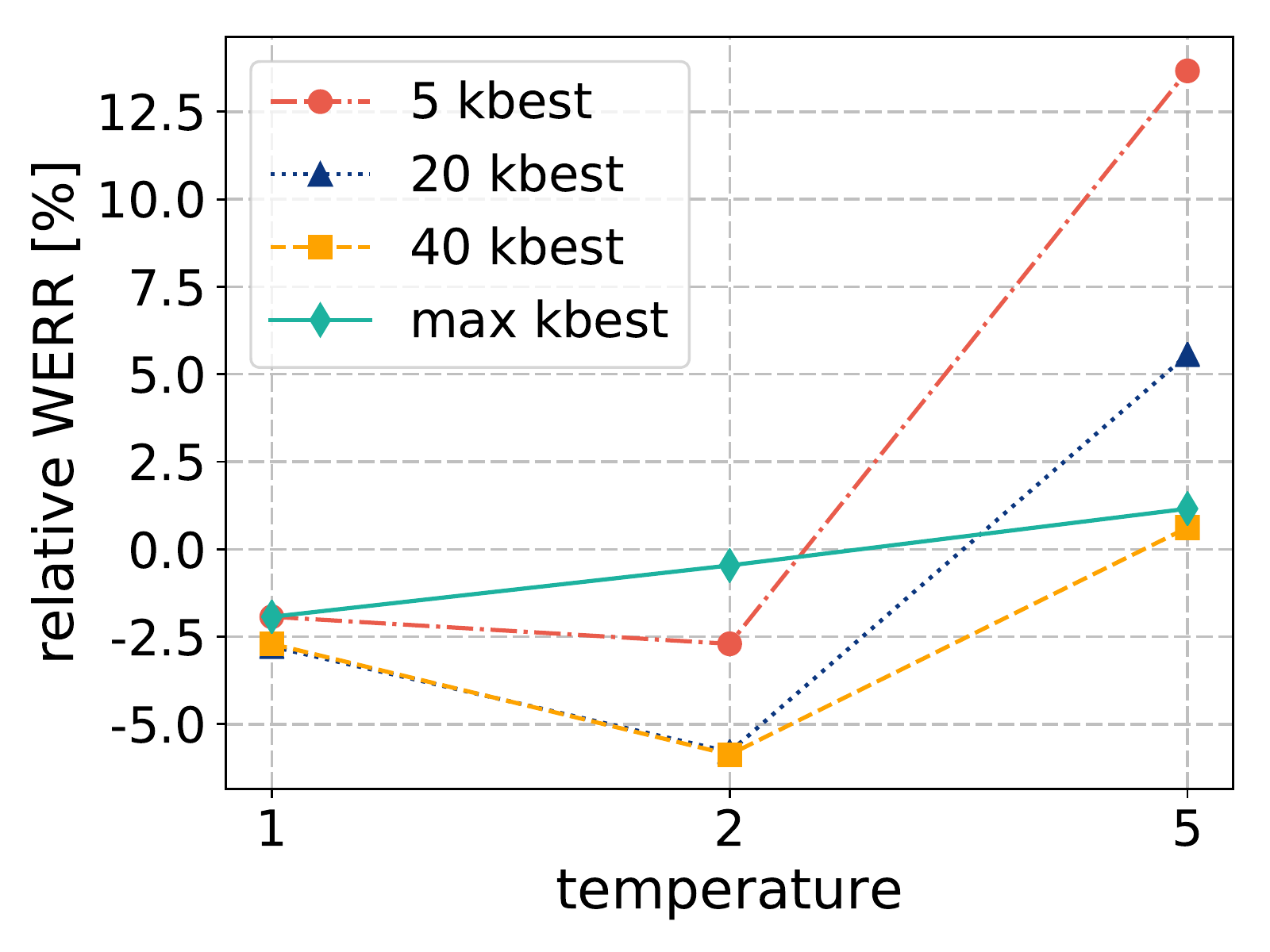}
    \textit{\footnotesize (a) clean test set}
  \end{minipage} %
  \begin{minipage}[t]{0.26\linewidth}
  \centering
    \includegraphics[width=\linewidth]{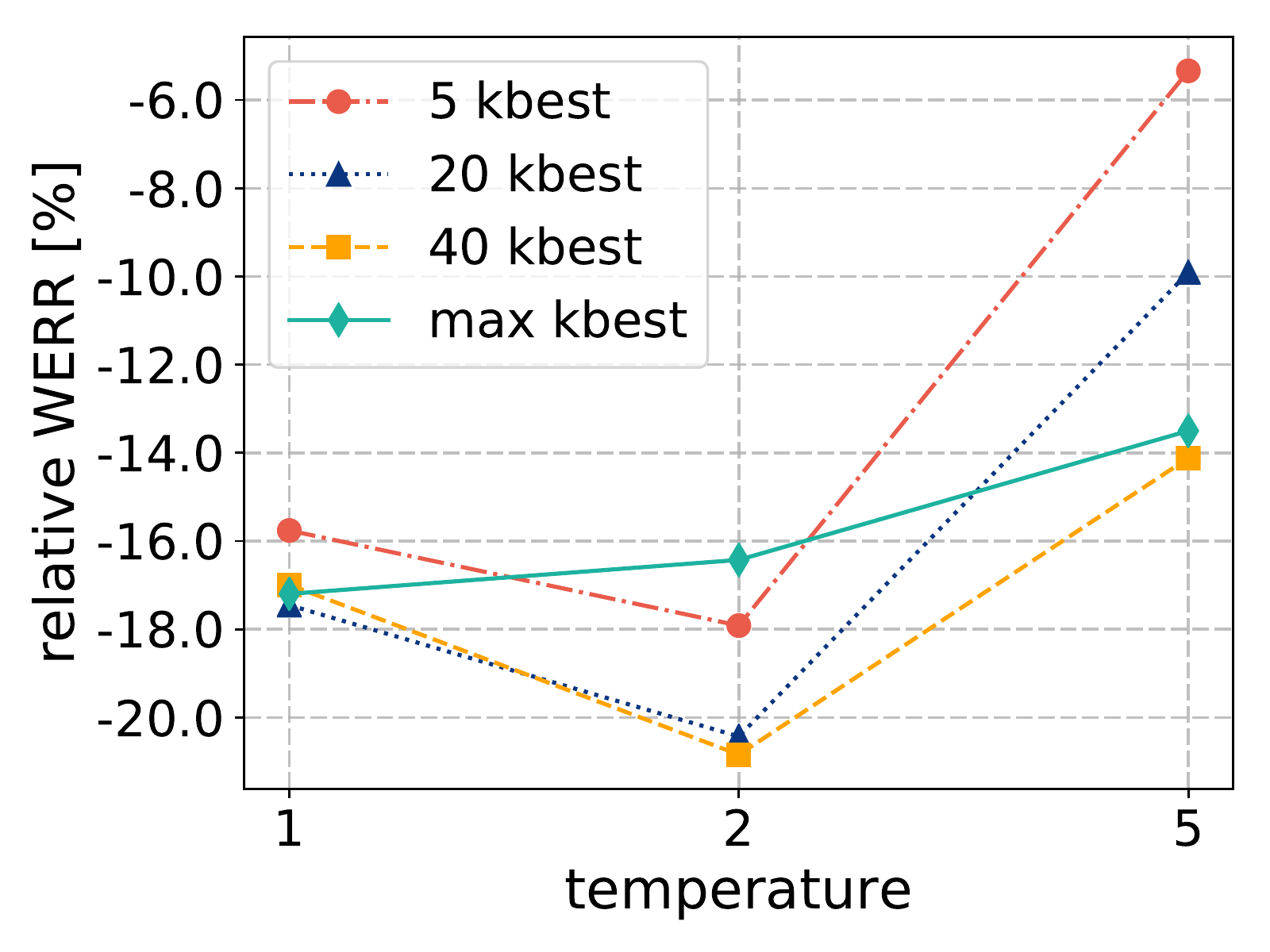}
    \textit{\footnotesize (b) noisy test set}
  \end{minipage} %
  \begin{minipage}[t]{0.26\linewidth}
  \centering
     \includegraphics[width=\linewidth]{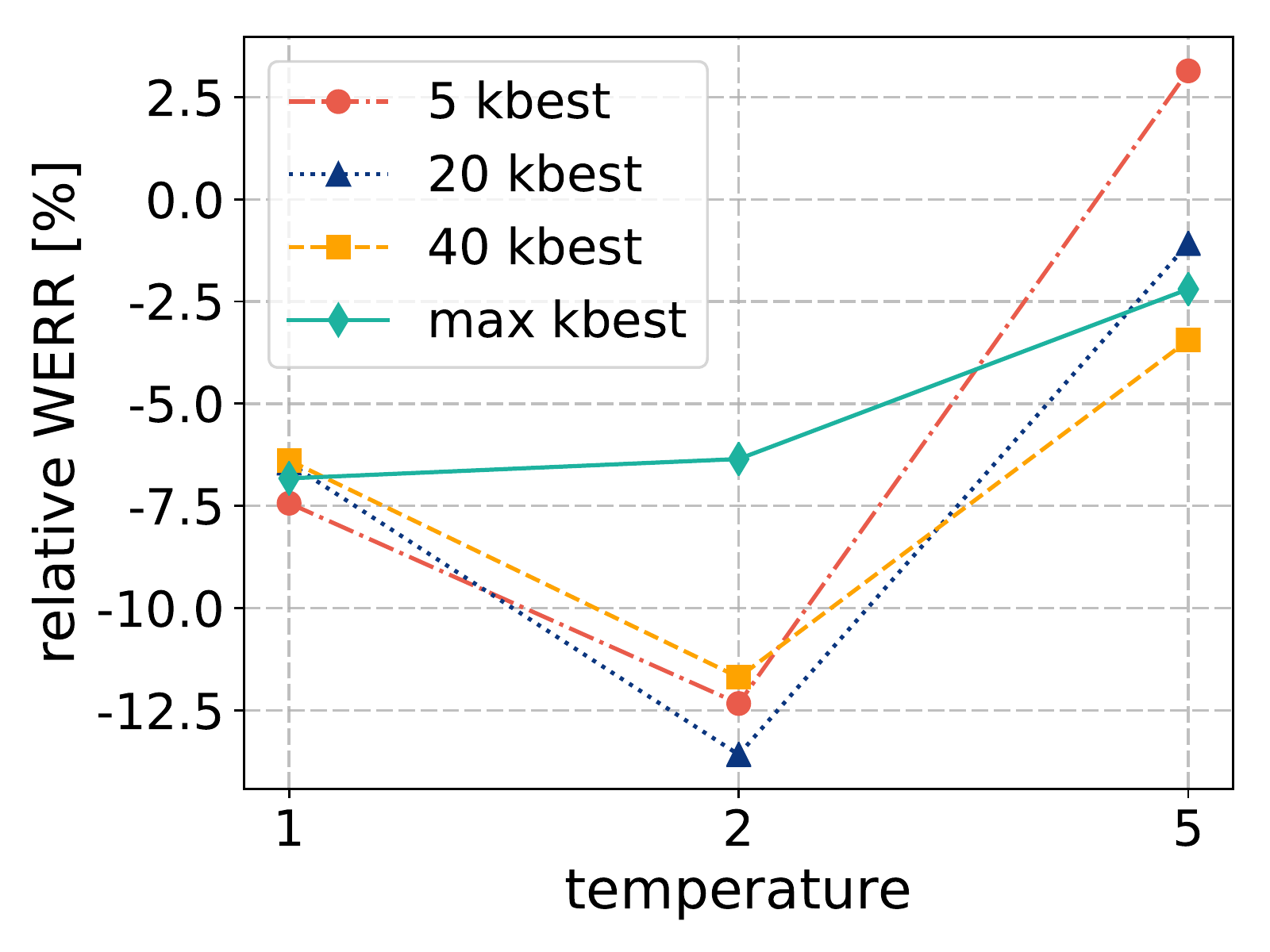}
     \textit{\footnotesize (c) realistic test set}
  \end{minipage} %
  \vspace{-0.3em}
  \caption{\small \textit{Relative word error rate reduction (WERR) [\%] (with respect to the cross-entropy trained teacher) from different student models when tunning the distillation temperature and logits selection. max kbest means preserving all logits.}}
  \vspace{-0.2em}
  \label{fig:encoding-and-temperature}
  \end{center}
  \vspace{-4mm}
\end{figure*}

We explore the resulting recognition accuracy for multiple settings of temperature $T$ and candidate logits selection $k$, which are summarized in Figure \ref{fig:encoding-and-temperature}.
\subsubsection{Temperature $T$ = 1}
While using a temperature of 1, no significant differences in WER can be seen even for a very aggressive logits selection (5 senones out of 3010). Inspecting the average output distribution after application of softmax, we find out that the highest probability is close to one and the rest close to zero. Therefore, a sum of probabilities of senones that do not belong to the 5-best is still small. Redistribution of this mass among 5-best senones then does not affect the distribution much.
\vspace{-3mm}
\subsubsection{Temperature $T$ = 2}
Interestingly, the combination of temperature 2 with logits selection bring accuracy improvements. We observe that taking 5-best values into consideration is not sufficient. However, the difference between results obtained using 20-best and 40-best logits is minimal.
\vspace{-3mm}
\subsubsection{Temperature $T$ = 5}
When the output distribution gets flatter (temperature 5), the difference among highest probabilities diminishes. Then the effect of multiplication by constant is similar for all $k$ senones and it becomes more difficult to estimate the correct one. The fewer candidates are taken into account, the more severe degradation occurs.\\

Based on the analysis, the most promising hyperparameters for our student models are: temperature $T= 2.0$ and $k=20$ for logits selection. These values are fixed for the following experiments.

\subsection{Size of the training dataset}
\label{sec:dataset-size}

\begin{figure*}[!h]
\begin{center}
\begin{minipage}[t]{0.26\linewidth}
  \centering
    \includegraphics[width=\linewidth]{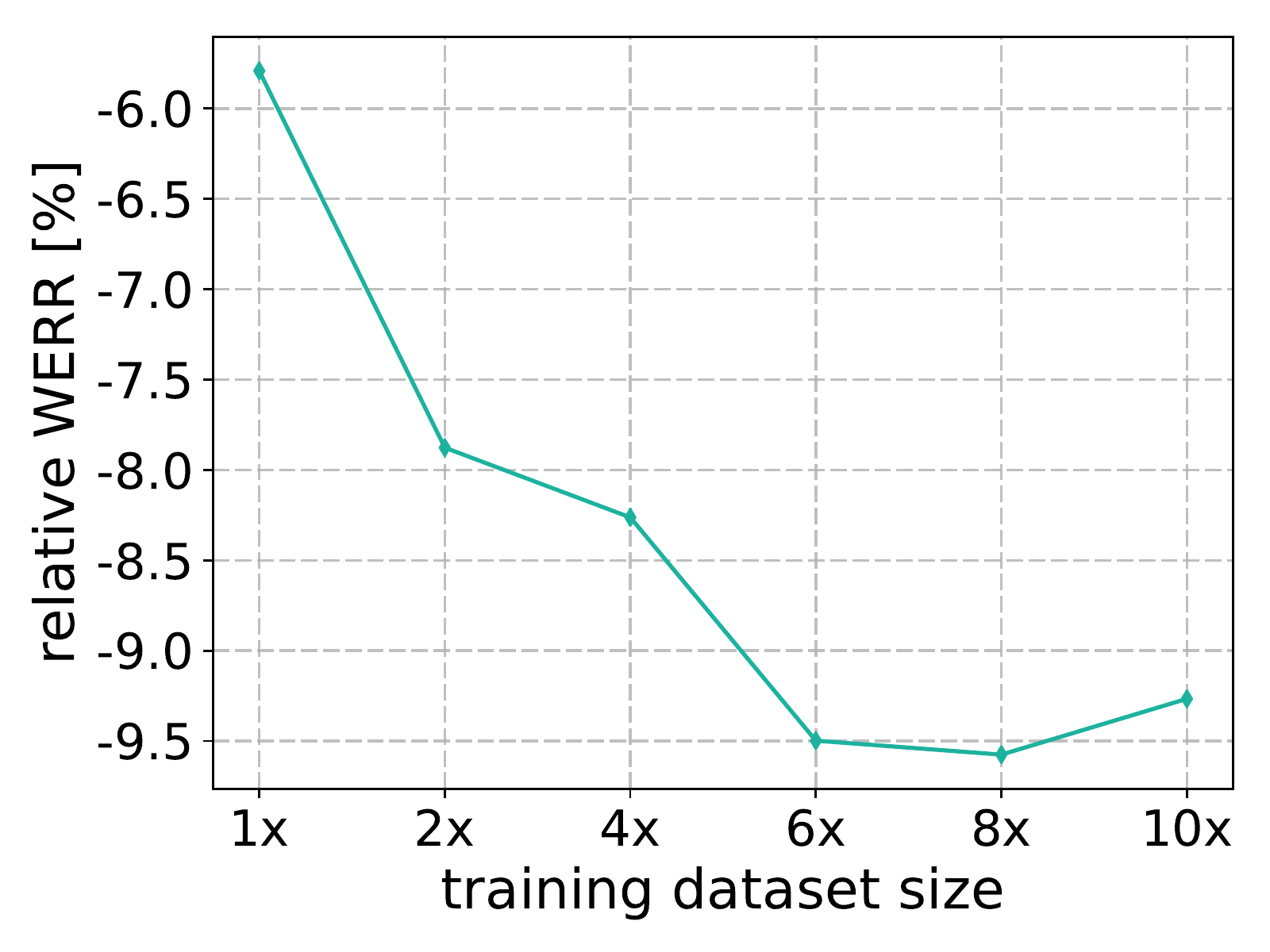}
    \textit{\footnotesize (a) clean test set}
  \end{minipage} %
  \begin{minipage}[t]{0.26\linewidth}
  \centering
    \includegraphics[width=\linewidth]{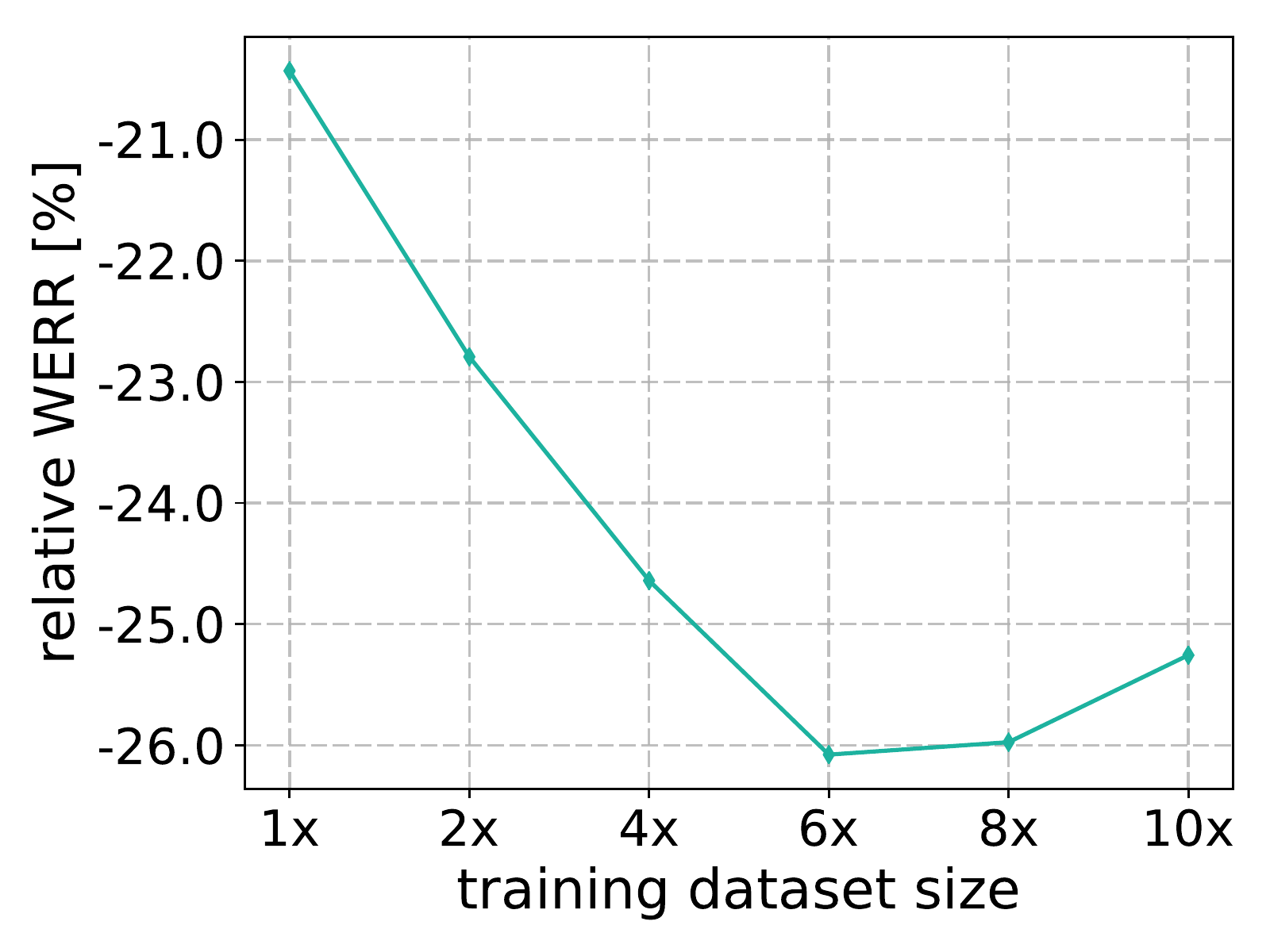}
    \textit{\footnotesize (b) noisy test set}
  \end{minipage} %
  \begin{minipage}[t]{0.26\linewidth}
  \centering
     \includegraphics[width=\linewidth]{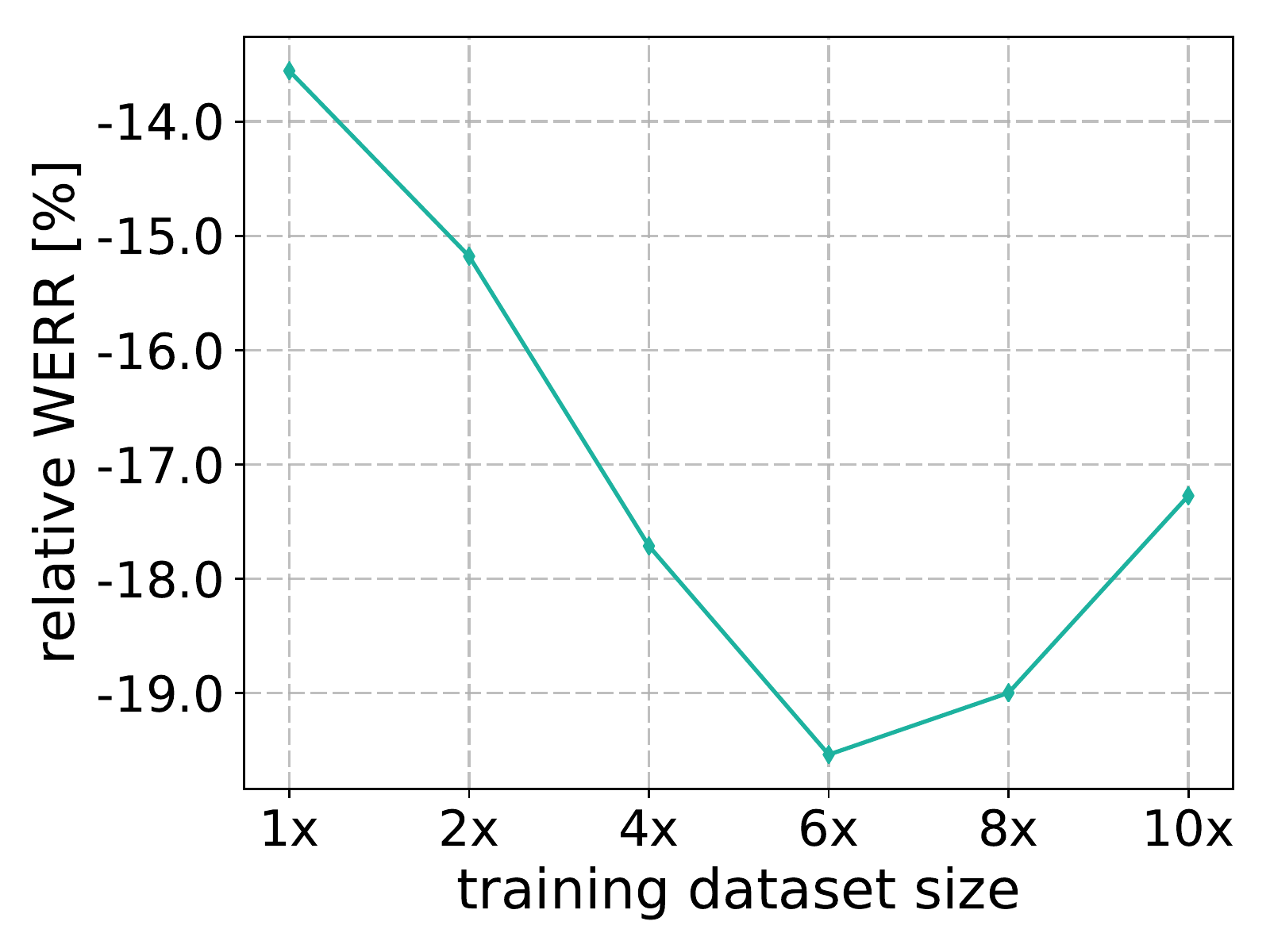}
     \textit{\footnotesize (c) realistic test set}
  \end{minipage} %
  \vspace{-0.3em}
  \caption{\small \textit{Relative word error rate reduction (WERR) [\%] (with respect to the cross-entropy trained teacher) from different student models incorporating more untranscribed training data. ``$1\times$" means $1 \times 800$ hours of training data. Fix $T=2, k=20$ for building all the models.}}
  \vspace{-0.2em}
  \label{fig:train-data-size}
  \end{center}
  \vspace{-4mm}
\end{figure*}

\begin{table*}[!h]\small
    \caption{\small \textit{Effects of applying sequence training on both teacher and student models. Results are expressed in terms of relative word error rate reduction (WERR) [\%] (with respect to the cross-entropy trained teacher). Minus sign indicates improvement. Relative WERR inside the bracket is computed with respect to the sequence trained teacher. $T=2, k=20$ for building all the student models.}}
    \centering
    \begin{tabular}{l l l l l l}
        \toprule
        Acoustic model & Training objective & Training data & Clean test set& Noisy test set& Realistic test set\\
        \midrule
        Teacher & xent & 800h clean &\hspace{2.8mm}0.00 &\hspace{1.7mm} 0.00 & \hspace{2.7mm}0.00 \\
        ~~Student & xent & $6\times800$h parallel & \hspace{1mm} -9.50 & -26.08 & -19.57 \\
        \midrule
        Teacher & xent, sMBR & 800h clean, 800h clean & \hspace{1.7mm}-5.79 \hspace{2mm} \textit{(0.00)}& \hspace{1.7mm}-0.98 \hspace{2mm} \textit{(0.00)}& \hspace{1.7mm}-3.68 \hspace{2mm} \textit{(0.00)}\\
        ~~Student & xent & $6\times800$h parallel& -11.89 \hspace{1mm} \textit{(-6.48)}   & -26.80  \textit{(-26.08)} & -21.63  \textit{(-18.63)}\\
        ~~Student & xent, sMBR & $6\times800$h parallel, 800h noisy& -15.29  \textit{\textbf{(-10.08)}} & -29.36 \textit{\textbf{(-28.67)}}& -22.54 \textit{\textbf{(-19.58)}}\\
        \bottomrule
    \end{tabular}
    \label{tab:seq-training}
\end{table*}
Building both the baseline and the multi-condition system require transcribed data. Compared to those methods, the major advantage of the teacher-student training approach is that it does not  require transcripts once the teacher model is trained. Therefore, we could explore how much the WER can be further reduced by incorporating even larger number of utterances in the training set to build the student model. 
The student training dataset is gradually increased up to ten times more audio compared with the original amount.
The relative word error rate reduction (WERR) is displayed in Figure \ref{fig:train-data-size} as a function of training data size. Similar trends are observed for all the test datasets. As expected, the accuracy improves with the increasing amount of data. However, the minimum WER is achieved when using approximately 4800 hours ($6\times$). It could be possible that we only have about 4800 hours of unique noise resources and the model overfits after adding repeated noise examples. This hypothesis, however, requires further investigation. Alternatively, the fact that the teacher itself is erroneous could also affect dependence of accuracy on training data size.

\subsection{Sequence training}
\label{sec:seq-training}
\vspace{-5mm}
Sequence training has been shown to be effective in improving ASR performance in general \cite{Vesely:Seq-training}. Wong and Gales also investigated the usefulness of the combination of sequence and teacher-student training \cite{Wong:Sequence-TS}. In our experiment, we at first fine-tune the original cross-entropy trained teacher using state-level minimum Bayes risk (sMBR) criterion \cite{gibson2006hypothesis}. As displayed in Table \ref{tab:seq-training}, the new teacher outperforms the original one on all our test datasets. We then train a new student network on top of the new sequence-trained teacher using the parallel corpus.  We used  size $6\times$ of the parallel datasets to train the student as it was shown to be the best option in the previous experiment. The new cross-entropy trained student network is able to make use of the improved teacher, since its performance is better than that of the student taught by the weaker teacher. Finally, the new student is further optimized by means of sMBR training using only the transcribed portion of noisy data ($1\times$).

\section{Conclusion}
\label{sec:conclusion}

In this paper, we explore the teacher-student learning approach using parallel clean and noisy corpus to improve speech recognition performance under multimedia noise. We gradually optimize this system by applying logits selection and incorporating larger amount of untranscribed training data. With a temperature of $T=2$ and logits selection of $k=20$ highest values, we obtain the best student model using a parallel clean and noisy corpus which is about 6 times more of the original clean training data. By applying standard sequence training on both the teacher and student model, the final student brings relative WER reductions of about 10.1\%, 28.7\% and 19.6\% on the clean, simulated noisy and real test sets, respectively.

In the future, a larger corpus of noise sources will be collected to prevent the multimedia samples from being repeated. We will attempt to scale up architecture and training datasets.
Soft targets selection based on teacher's certainty could be explored as well.

\newpage
\bibliographystyle{IEEEbib}
\footnotesize
\bibliography{refs}

\end{document}